\documentclass[a4paper,12pt]{article}
\usepackage{times, url}
\usepackage{graphicx,amsmath,amssymb,amsthm,amsfonts,bm,float,cite}

\usepackage[none]{hyphenat}
\usepackage{ragged2e}
\justifying
\usepackage[center]{caption}
\begin{document}
\setcounter{page}{1}

\thispagestyle{empty}

\begin{center}
{\LARGE \bf  Reverse Degree-Based Topological Indices and QSPR Analysis of Cancer Drugs \\[4mm] } 
\vspace{8mm}

{\Large \bf H. M. Nagesh}
\vspace{3mm}

Department of Science and Humanities, \\
PES University, Bangalore, India. \\
e-mail: \url{hmnagesh1982@gmail.com}

\end{center}

\noindent

{\bf Abstract:} A topological index of a graph $G$ is a numerical quantity that describes its topology. Reverse degree-based topological indices play an important role in finding topological descriptors. Azacitidine, Decitabine, and Guadecitabine are hypomethylating agents which are used for the treatment of patients with higher-risk myelodyplastic syndromes, acute myeloid leukemia, and chronic myelomonocytic leukemia which are not suitable for in-depth treatments such as induction chemotherapy. In this article, some reverse degree-based topological indices of the three said drugs are computed. Furthermore, QSPR analysis of the said topological indices is discussed and it is shown that these topological indices are highly correlated with the physical properties of the three cancer drugs. These findings may help chemists and people working in the pharmaceutical industry to predict the properties of cancer drugs without experimenting.  \\\\
{\bf Keywords:} Reverse degree; reverse topological indices; Azacitidine; Decitabine; Guadecitabine. \\ 
\vspace{2mm}
\section{Introduction} \label{sec:Intr}
Indeed, the rapid advancements in technology and chemical/pharmaceutical techniques have led to the discovery of numerous new nano-materials, crystalline materials, and drugs. However, the vast number of new compounds and drugs generated each year poses a significant challenge for chemical and pharmaceutical researchers. Traditional methods of determining the chemical properties of these compounds through experimental means can be time-consuming. Fortunately, researchers have observed a strong correlation between the topological molecular structures of compounds and their physical behaviors, chemical characteristics, and biological features. This correlation has been found to influence properties, such as melting point, boiling point, and toxicity of drugs \cite{1,2,3,4,5}. 
H. Wiener \cite{6} discussed the development and application of quantitative structure-property relationship (QSPR) models. 

These models are computational tools that use mathematical algorithms to establish relationships between molecular structures and their corresponding properties or behaviors. QSPR models can predict various chemical and physical properties of compounds based on their molecular structures, thereby reducing the reliance on extensive experimental testing. By leveraging these QSPR models, researchers can effectively analyze and predict the properties of compounds and drugs without the need for exhaustive experimental testing. This approach significantly reduces the workload of chemical and pharmaceutical researchers, accelerates the drug discovery process, and facilitates the design of new materials with desired properties. 

The topological index of a molecule is a non-empirical numerical quantity that characterizes the structural features and branching patterns of the molecule. Topological indices are important tools for investigating many physicochemical properties of molecules without performing any testing. Many types of topological indices of graphs are classified into distance-based topological indices, degree-based topological indices, and spectrum-based topological indices. Among these, degree-based topological indices play a vital role in theoretical chemistry and pharmacology. Some important degree-based indices are the Randi\'c index, Zagreb indices, Harmonic index, sum connectivity index, etc. For example, the Randi\'c index is one of the excellent molecular descriptors in QSAR (Quantitative Structure-Activity Relationship) studies and is desirable for measuring the extent of branching of the carbon-atom skeleton of saturated hydrocarbons \cite{7,8,9,10,11}. 

The drugs Azacitidine, Decitabine, and Guadecitabine are important in the treatment of various hematological disorders, that is, myelodysplastic syndromes (MDS) and acute myeloid leukemia (AML). These drugs belong to a class of medications known as hypomethylating agents, which play a crucial role in altering gene expression by inhibiting DNA methyltransferases and offer significant advancements in the treatment of hematological disorders, particularly for patients with MDS and AML. By targeting DNA methyltransferases and modifying gene expression, they can potentially slow down disease progression and improve patient outcomes. For more details on Azacitidine, Decitabine, and Guadecitabine, the readers are referred to \cite{12,13}.

Let $G=(V,E)$ be a graph with vertex set $V(G)$ and edge set $E(G)$. The \emph{degree} of a vertex $v$ in $G$ is the number of vertices adjacent to $v$. It is common to denote the degree of a vertex $v$ by $d(v)$. The maximum and minimum degree of $G$ is denoted by $\Delta(G)$ and $\delta(G)$, respectively.

The concept of reverse vertex degree in a graph $G$ was introduced by Kulli in \cite{14}. The \emph{reverse vertex degree} in $G$ is defined by $\mathcal{R}_{\psi}(G)=\Delta(G)-d(v)+1$. Zhao et al. \cite{15} computed some of the reverse degree-based topological indices, such as, reverse general Randi\'c index, reverse Balaban index, reverse atom bond connectivity index, reverse geometric index, reverse Zagreb type indices, and reverse augmented Zagreb index for metal-organic networks TM-TCNB. Jung et al. \cite{16} computed the first and second reverse Zagreb indices, first and second reverse hyper Zagreb indices, reverse atomic-bond connectivity index and reverse geometric-arithmetic index for TUC4[m,n]. 

Wei et al. \cite{17} computed some reverse topological indices, namely, the reverse general Rand\'c index, reverse atom bond connectivity index, reverse geometric arithmetic index, reverse forgotten index, reverse Balaban index, and reverse Zagreb type indices for Remdisivir compound used in the treatment of Coronavirus (COVID 19). Ravi et al. \cite{18} analyzed the QSPR of the reverse degree-based topological indices. \newpage Haoer and Virk \cite{19} computed reverse Zagreb indices, reverse hyper indices, and their polynomials for metal-organic networks. Hashmi et al. \cite{20} computed the reverse ABC index for different generations of dendrimers. Prasanna Poojari et al. \cite{21} computed some reverse degree-based topological indices for Zanamivir and Oseltamivir. 

Monjit Chamua et al. \cite{22} computed the M-polynomial and neighborhood M-polynomial of some concise drug structures used in the treatment of cancer, that is, Azacitidine, Decitabine, and Guadecitabine. Furthermore, using these polynomials, some degree-based and neighborhood degree sum-based topological indices were computed. However, studies on the QSPR analysis of these cancer drugs were not attempted. In this paper, an attempt is made to fill this gap and compute some reverse degree-based topological indices of the molecular structures of Azacitidine, Decitabine, and Guadecitabine to study the QSPR analysis. 

In the QSPR study, physicochemical properties and topological indices are used to predict the bioactivity of the chemical compounds. In this regard, the physical properties of Azacitidine, Decitabine, and Guadecitabine, such as boiling point (BP), enthalpy of vaporization (E), flash point (P), molar refraction (MR), polar surface area (PSA), polarizability (P), surface tension (T), and molar volume (MV) are considered. Some reverse degree-based topological indices of the molecular structures of Azacitidine, Decitabine, and Guadecitabine are computed.    

\subsection{Basic definitions}  

Milan Randi\'c \cite{23} introduced the first degree-based index. Wei et al. \cite{17}  defined the reverse Randi\'c index as:  
\begin{equation}
\mathcal{R}R_{\alpha}(G)=\displaystyle \sum_{mn \in E(G)} \left(R_{\psi(m)} \times R_{\psi}(n) \right)^{\alpha}; \alpha=1,-1,\frac{1}{2}, -\frac{1}{2} 
\end{equation} 
If $\alpha =1$, then it is called the \emph{reverse second Zagreb index} $\mathcal{R}M_{2}(G)$. 

Estrada et al. \cite{24} presented the atom bond connectivity index. Wei et al.\cite{17} defined the reverse atom bond connectivity index as:
\begin{equation}
\mathcal{R}ABC(G)=\displaystyle \sum_{mn \in E(G)} \sqrt{\frac{R_{\psi(m)} + R_{\psi}(n)-2}{R_{\psi(m)} \times R_{\psi}(n)}}
\end{equation} 
Vukicevic et al. \cite{25} proposed the geometric arithmetic index. Wei et al. \cite{17} defined the reverse geometric arithmetic index as: 
\begin{equation}
\mathcal{R}GA(G)=\displaystyle \sum_{mn \in E(G)} 2 \frac{\sqrt{R_{\psi(m)} \times R_{\psi}(n)}}{R_{\psi(m)} + R_{\psi}(n)} 
\end{equation} 
In 1972, Gutman discussed the first and second Zagreb indices \cite{26}. Wei et al. \cite{17} defined the reverse first and reverse second Zagreb indices as:
\begin{equation}
\mathcal{R}M_{1}(G)=\displaystyle \sum_{mn \in E(G)} \left(R_{\psi(m)} + R_{\psi(n)}\right) 
\end{equation}
\begin{equation}
\mathcal{R}M_{2}(G)=\displaystyle \sum_{mn \in E(G)} \left(R_{\psi(m)} \times R_{\psi(n)}\right) 
\end{equation}
In 2008, Doslic and Gutman \cite{27,28} presented the first and second Zagreb co-indices. Wei et al. \cite{17} defined the reverse first and reverse second Zagreb co-indices as:
\begin{equation}
\mathcal{R}\overline{M_{1}}(G)=2|E(G)|\left(|(V(G)|-1 \right)-\mathcal{R}M_{1}(G)
\end{equation}
\begin{equation}
\mathcal{R}\overline{M_{2}}(G)=2|E(G)|^{2}-\frac{1}{2} \mathcal{R}M_{1}(G)-\mathcal{R}M_{2}(G)
\end{equation}
In 2013, Shirdel et al. \cite{29} discussed the concept of hyper Zagreb index. Wei et al. \cite{17} defined the reverse hyper Zagreb index as:
\begin{equation}
\mathcal{R}HM(G)=\displaystyle \sum_{mn \in E(G)}  \left(R_{\psi(m)} + R_{\psi(n)}\right)^2
\end{equation}
Furtula and Gutman \cite{30} introduced the notion of Forgotten index. Wei et al. \cite{17} defined the reverse forgotten index as:
\begin{equation}
\mathcal{R}F(G)=\displaystyle \sum_{mn \in E(G)} \left((R_{\psi(m)})^2 + (R_{\psi(n)})^2\right)
\end{equation}
Wei et al. \cite{17} defined the reverse Balaban index for a graph as follows: Let $n$ and $m$ be order and size of a graph $G$, respectively. Then,
\begin{equation}
\mathcal{R}J(G)=\frac{m}{m-n+2} \displaystyle \sum_{mn \in E(G)} \frac{1}{\sqrt{R_{\psi(m)} \times R_{\psi(n)}}}
\end{equation}
Wei et al. \cite{17} defined the reverse first multiple and reverse second multiple Zagreb indices as:
\begin{equation}
\mathcal{R}PM_{1}(G)=\displaystyle \prod_{mn \in E(G)} \left(R_{\psi(m)} + R_{\psi(n)} \right) 
\end{equation}
\begin{equation}
\mathcal{R}PM_{2}(G)=\displaystyle \prod_{mn \in E(G)} \left(R_{\psi(m)} \times R_{\psi(n)} \right) 
\end{equation}
Rajini et al. \cite{31} introduced the concept of redefined first, second and third Zagreb indices for a graph. Wei et al. \cite{7} defined the reverse redefined first, second, and third Zagreb indices as:
\begin{equation}
\mathcal{R}ReZG_{1}(G)=\displaystyle \sum_{mn \in E(G)} \frac{R_{\psi(m)} + R_{\psi(n)}}{R_{\psi(m)} \times R_{\psi(n)}}
\end{equation}
\begin{equation}
\mathcal{R}ReZG_{2}(G)=\displaystyle \sum_{mn \in E(G)} \frac{R_{\psi(m)} \times R_{\psi(n)}}{R_{\psi(m)} + R_{\psi(n)}}
\end{equation}
\begin{equation}
\mathcal{R}ReZG_{3}(G)=\displaystyle \sum_{mn \in E(G)} \left(R_{\psi(m)} + R_{\psi(n)}\right)\left(R_{\psi(m)} \times R_{\psi(n)} \right)
\end{equation}
\section{Methods and techniques}
The methods used in this paper contain reverse vertex degree counting, division of vertices based on reverse degree, and partition of edges based on the reverse degree of end vertices. The topological indices as given in formulas (1-15) are calculated with the help of reverse vertex degree counting and partition of edges techniques. The JMP software is used for finding correlation coefficients. The chemical structure (both 2D and 3D) of Azacitidine, Decitabine, and Guadecitabine are taken from PubChem, and molecular graphs of chemical structures are drawn using Microsoft Word. There are eight physicochemical properties of cancer drugs under consideration for the analysis. The properties are boiling point (BP), enthalpy of vaporization (E), flash point (P), molar refraction (MR), polar surface area (PSA), polarizability (P), surface tension (T), and molar volume (MV). These properties are shown in Table 4 and are collected from ChemSpider. 

\section{Reverse topological indices of Azacitidine}
In this section, we first find the reverse degree-based topological indices of Azacitidine and then present the graphical comparison of topological indices. The chemical structure and molecular graph of Azacitidine are shown in Figure 1.   
\vspace{5mm}
\begin{figure}[h!]
\centering
\includegraphics[width=100mm]{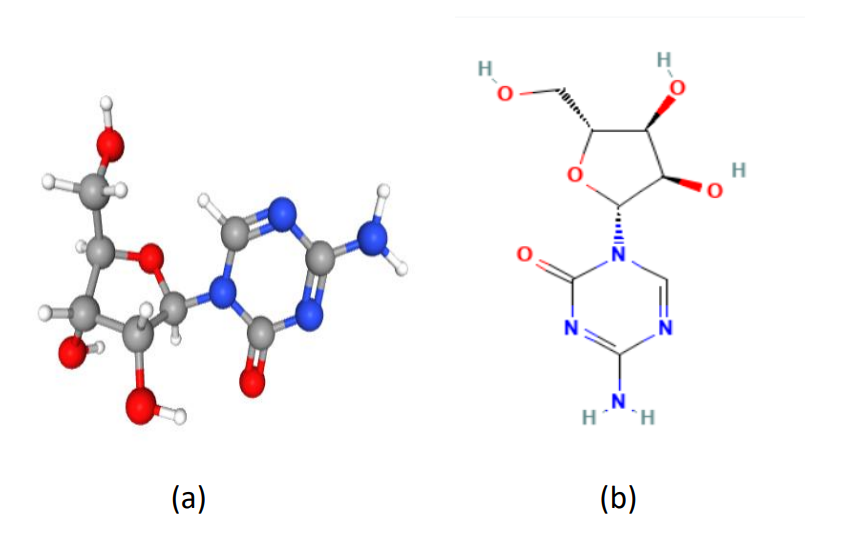}
  \caption{(a) Chemical structure of Azacitidine. (b) molecular graph of Azacitidine. }
  \end{figure} 
  
We consider the hydrogen-suppressed molecular graph of compounds since the vertices representing hydrogen atoms do not contribute graph isomorphism. The graph of Azacitidine with vertices and edges is shown in Figure 2. 
\vspace{5mm}
\begin{figure}[h!]
\centering
\includegraphics[width=45mm]{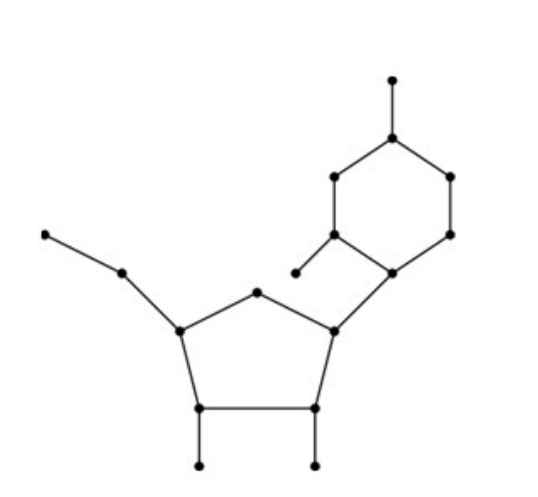}
  \caption{Molecular graph of Azacitidine with vertices and edges.}
  \end{figure} 
\newpage
The number of vertices and edges of the molecular graph of Azacitidine are 17 and 18, respectively. The edge set of Azacitidine is partitioned into five sets based on the reverse degree of end vertices. 

The first edge set of the partition includes 5 edges $mn$, where $R_{\psi(m)}=1$ and $R_{\psi(n)}=1$. The second edge set of the partition includes 7 edges $mn$, where $R_{\psi(m)}=1$ and $R_{\psi(n)}=2$. The third edge set of the partition includes 4 edges $mn$, where $R_{\psi(m)}=1$ and $R_{\psi(n)}=3$. The fourth edge set of the partition includes an edge $mn$, where $R_{\psi(m)}=2$ and $R_{\psi(n)}=2$. The fifth edge set of the partition includes an edge $mn$, where $R_{\psi(m)}=2$ and $R_{\psi(n)}=3$.   

The maximum degree of the molecular graph of Azacitidine is 3. By using the definition of reverse vertex degree $R_{\psi}(G)=\Delta(G)-d(v)+1$, the reverse degree-based edge partition of Azacitidine is given in Table 1. 

\begin{center}
\begin{tabular}{||c c c c c c ||} 
 \hline
 $\left(R_{\psi(m)}, R_{\psi(m)}\right)$ & (1,1) & (1,2) & (1,3) & (2,2) & (2,3) \\ [0.5ex] 
 \hline\hline
 Number of edges & 5 & 7 & 4 & 1 & 1 \\ 
 \hline 
\end{tabular}
\end{center}
\begin{center}
\textbf{Table 1}. Edge partition of Azacitidine.
\end{center}  
We now find reverse degree-based topological indices of Azacitidine.\\\\
$\bullet$ Reverse Randi\'c index
\begin{equation*}
\mathcal{R}R_{\alpha}(G)=\displaystyle \sum_{mn \in E(G)} \left(R_{\psi(m)} \times R_{\psi}(n) \right)^{\alpha}; \alpha=1,-1,\frac{1}{2}, -\frac{1}{2} 
\end{equation*} 
For $\alpha=1$,
\begin{align*}
\mathcal{R}R_{1}(G)=& 5(1 \times 1)+7(1 \times 2)+4(1 \times 3)+1(2 \times 2)+1(2 \times 3)=41.
\end{align*}
For $\alpha=-1$, $\mathcal{R}R_{-1}(G)=\frac{5}{1 \times 1}+\frac{7}{1 \times 2}+\frac{4}{1 \times 3}+\frac{1}{2 \times 2}+\frac{1}{2 \times 3}=10.25$.\\
For $\alpha=\frac{1}{2}$, 
\begin{align*}
\mathcal{R}R_{\frac{1}{2}}(G)=&5\sqrt{1 \times 1}+7\sqrt{1 \times 2}+4\sqrt{1 \times 3}+1\sqrt{2 \times 2}+1\sqrt{2 \times 3}=26.28.
\end{align*}
For $\alpha=-\frac{1}{2}$, 
\begin{align*}
\mathcal{R}R_{-\frac{1}{2}}(G)=&\frac{5}{\sqrt{1 \times 1}}+\frac{7}{\sqrt{1 \times 2}}+\frac{4}{\sqrt{1 \times 3}}+\frac{1}{\sqrt{2 \times 2}}+\frac{1}{\sqrt{2 \times 3}}=13.17.
\end{align*}
$\bullet$ Reverse atom bond connectivity index
\begin{align*}
\mathcal{R}ABC(G) & = \displaystyle \sum_{mn \in E(G)} \sqrt{\frac{R_{\psi(m)} + R_{\psi}(n)-2}{R_{\psi(m)} \times R_{\psi}(n)}} \\
& =5\sqrt{\frac{1+1-2}{1 \times 1}}+ 7\sqrt{\frac{1+2-2}{1 \times 2}}+4\sqrt{\frac{1+3-2}{1 \times 3}}+1\sqrt{\frac{2+2-2}{2 \times 2}}+1\sqrt{\frac{2+3-2}{2 \times 3}} \\
& = 9.63.
\end{align*} 
$\bullet$ Reverse geometric arithmetic index
\begin{align*}
\mathcal{R}GA(G)=& \displaystyle \sum_{mn \in E(G)} 2 \frac{\sqrt{R_{\psi(m)} \times R_{\psi}(n)}}{R_{\psi(m)} + R_{\psi}(n)} \\
& = 5\left(\frac{2\sqrt{1 \times 1}}{1+1}\right)+7\left(\frac{2\sqrt{1 \times 2}}{1+2}\right)+4\left(\frac{2\sqrt{1 \times 3}}{1+3}\right) + 1\left(\frac{2\sqrt{2 \times 2}}{2+2}\right)+1\left(\frac{2\sqrt{2 \times 3}}{2+3}\right) \\
 & = 17.04.
\end{align*}
$\bullet$ Reverse first Zagreb index
\begin{align*}
\mathcal{R}M_{1}(G)=& \displaystyle \sum_{mn \in E(G)} \left( R_{\psi(m)} + R_{\psi(n)} \right)\\
& = 5(1+1)+7(1+2)+4(1+3)+1(2+2)+1(2+3)= 56.
\end{align*}
$\bullet$ Reverse second Zagreb index
\begin{align*}
\mathcal{R}M_{2}(G)=& \displaystyle \sum_{mn \in E(G)}  \left(R_{\psi(m)} \times R_{\psi(n)}\right) \\
& = 5(1 \times 1)+7(1\times2)+4(1\times3)+1(2\times2)+1(2\times3)=41.
\end{align*} 
$\bullet$ Reverse first Zagreb co-index
\begin{align*}
\mathcal{R}\overline{M_{1}}(G)=& 2|E(G)|\left(|(V(G)|-1 \right)-\mathcal{R}M_{1}(G) = 2(18)(17-1)-56 = 520.
\end{align*} 
$\bullet$ Reverse second Zagreb co-index
\begin{align*}
\mathcal{R}\overline{M_{2}}(G)=&2|E(G)|^{2}-\frac{1}{2} \mathcal{R}M_{1}(G)-\mathcal{R}M_{2}(G) = 2(18)^2-\frac{56}{2}-41 = 579.
\end{align*}
$\bullet$ Reverse hyper Zagreb index
\begin{align*}
\mathcal{R}HM(G)=& \displaystyle \sum_{mn \in E(G)}  \left(R_{\psi(m)} + R_{\psi(n)}\right)^2 \\
& = 5(1+1)^2+7(1+2)^2+4(1+3)^2+1(2+2)^2+1(2+3)^2= 188.
\end{align*}
$\bullet$ Reverse forgotten index
\begin{align*}
\mathcal{R}F(G)=& \displaystyle \sum_{mn \in E(G)} (R_{\psi(m)})^2 + (R_{\psi(n)})^2\\
& = 5(1^2+1^2)+7(1^2+2^2)+4(1^2+3^2)+1(2^2+2^2)+1(2^2+3^2)= 106.
\end{align*}
$\bullet$ Reverse Balaban index: Let $n$ and $m$ be order and size of $G$, respectively. Then
\begin{align*}
\mathcal{R}J(G)=& \frac{m}{m-n+2} \displaystyle \sum_{mn \in E(G)} \frac{1}{\sqrt{R_{\psi(m)} \times R_{\psi(n)}}}\\
& = \left(\frac{18}{18-17+2} \right)\left(\frac{5}{\sqrt{1 \times 1}}+\frac{7}{\sqrt{1 \times 2}}+\frac{4}{\sqrt{1 \times 3}}+\frac{1}{\sqrt{2 \times 2}}+\frac{1}{\sqrt{2 \times 3}}\right) = 79.004.
\end{align*}
$\bullet$ Reverse first multiple Zagreb index
\begin{align*}
\mathcal{R}PM_{1}(G)=&\displaystyle \prod_{mn \in E(G)} \left(R_{\psi(m)} + R_{\psi(n)} \right)\\
& = 5(1+1)\times 7(1+2)\times 4(1+3)\times 1(2+2)\times 1(2+3)= 67200.
\end{align*}
$\bullet$ Reverse second multiple Zagreb index
\begin{align*}
\mathcal{R}PM_{2}(G)=&\displaystyle \prod_{mn \in E(G)} \left(R_{\psi(m)} \times R_{\psi(n)} \right)\\
& = 5(1\times 1)\times 7(1\times 2)\times 4(1\times 3)\times 1(2\times 2)\times 1(2\times 3)= 20160.
\end{align*}
$\bullet$ Reverse first refined Zagreb index
\begin{align*}
\mathcal{R}ReZG_{1}(G)=&\displaystyle \sum_{mn \in E(G)} \frac{R_{\psi(m)} + R_{\psi(n)}}{R_{\psi(m)} \times R_{\psi(n)}}\\
& = 5\left(\frac{1+1}{1 \times 1} \right)+7\left(\frac{1+2}{1 \times 2} \right)+4\left(\frac{2+3}{2 \times 3} \right)+1\left(\frac{2+2}{2 \times 2} \right)+1\left(\frac{2+3}{2 \times 3} \right)= 27.67.
\end{align*}
$\bullet$ Reverse second refined Zagreb index
\begin{align*}
\mathcal{R}ReZG_{2}(G)=&\displaystyle \sum_{mn \in E(G)} \frac{R_{\psi(m)} \times R_{\psi(n)}}{R_{\psi(m)} + R_{\psi(n)}}\\
& = 5\left(\frac{1\times 1}{1 + 1} \right)+7\left(\frac{1\times 2}{1 + 2} \right)+4\left(\frac{2\times3}{2 + 3} \right)+1\left(\frac{2\times2}{2 + 2} \right)+1\left(\frac{2\times3}{2 + 3} \right)= 12.37.
\end{align*}
$\bullet$ Reverse third refined Zagreb index
\begin{align*}
\mathcal{R}ReZG_{3}(G)=&\displaystyle \sum_{mn \in E(G)} \left(R_{\psi(m)} + R_{\psi(n)}\right)\left(R_{\psi(m)} \times R_{\psi(n)} \right)\\
& = 5(2 \times 1)+7(3 \times 2)+4(4 \times 3)+1(4\times 4)+1(5 \times 6)= 146.
\end{align*}  
 
\section{Reverse topological indices of Decitabine}
We find the reverse degree-based topological indices of Decitabine and present the graphical comparison. The chemical structure and molecular graph of Decitabine are shown in Figure 3.   
\newpage
\vspace{5mm}
\begin{figure}[h!]
\centering
\includegraphics[width=120mm]{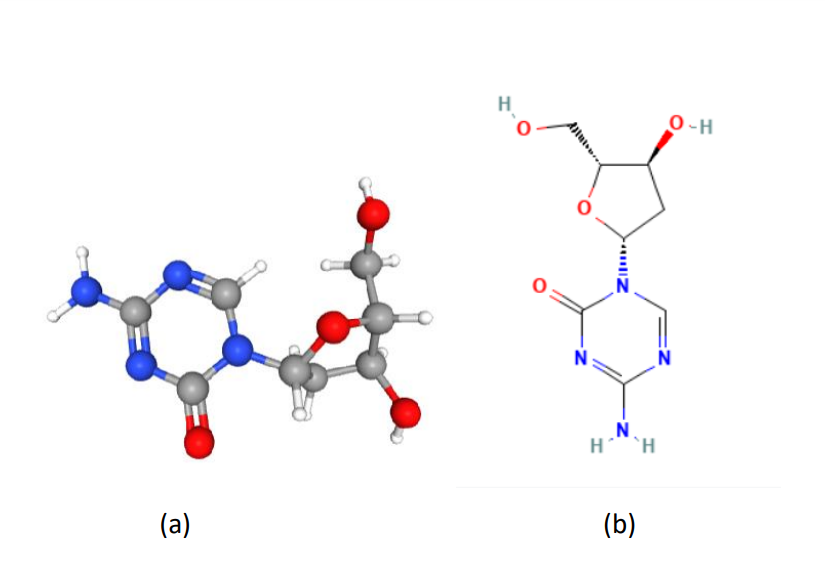}
  \caption{(a) Chemical structure of Decitabine. (b) molecular graph of Decitabine. }
  \end{figure}  
    
The graph of Decitabine with vertices and edges is shown in Figure 4. 
\vspace{5mm}
\begin{figure}[h!]
\centering
\includegraphics[width=80mm]{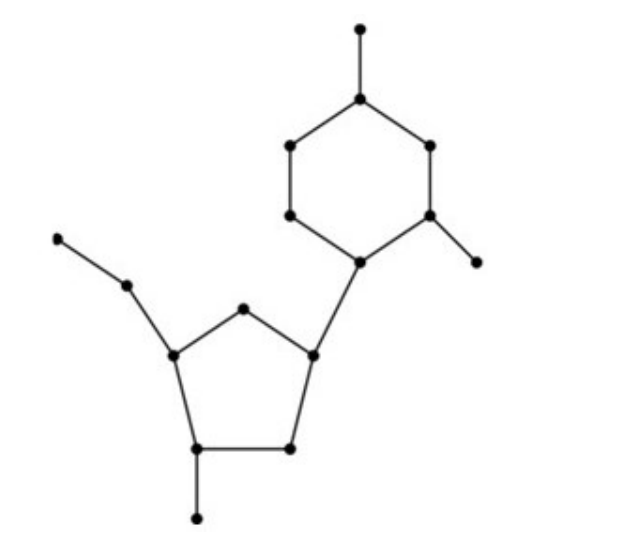}
  \caption{Molecular graph of Decitabine with vertices and edges. }
  \end{figure} 

The number of vertices and edges of the molecular graph of Decitabine are 16 and 17, respectively. The edge set of Decitabine is partitioned into five sets based on the reverse degree of end vertices. 

The first edge set of the partition includes 3 edges $mn$, where $R_{\psi(m)}=1$ and $R_{\psi(n)}=1$. The second edge set of the partition includes 9 edges $mn$, where $R_{\psi(m)}=1$ and $R_{\psi(n)}=2$. The third edge set of the partition includes 3 edges $mn$, where $R_{\psi(m)}=1$ and $R_{\psi(n)}=3$. The fourth edge set of the partition includes an edge $mn$, where $R_{\psi(m)}=2$ and $R_{\psi(n)}=2$. The fifth edge set of the partition includes an edge $mn$, where $R_{\psi(m)}=2$ and $R_{\psi(n)}=3$. 

The maximum degree of the molecular graph of Decitabine is 3. The reverse degree-based edge partition of Decitabine is given in Table 2. 

\begin{center}
\begin{tabular}{||c c c c c c ||} 
 \hline
 $\left(R_{\psi(m)}, R_{\psi(m)}\right)$ & (1,1) & (1,2) & (1,3) & (2,2) & (2,3)  \\ [0.5ex] 
 \hline\hline
 Number of edges & 3 & 9 & 3 & 1 & 1  \\ 
 \hline 
\end{tabular}
\end{center}
\begin{center}
\textbf{Table 2}. Edge partition of Decitabine.
\end{center}  
We now find obtain the reverse degree-based topological indices of Decitabine.\\
$\bullet$ Reverse Randi\'c index
\begin{equation*}
\mathcal{R}R_{\alpha}(G)=\displaystyle \sum_{mn \in E(G)} \left(R_{\psi(m)} \times R_{\psi}(n) \right)^{\alpha}; \alpha=1,-1,\frac{1}{2}, -\frac{1}{2} 
\end{equation*} 
For $\alpha=1$,
\begin{align*}
\mathcal{R}R_{1}(G)=& 3(1 \times 1)+9(1 \times 2)+3(1 \times 3)+1(2 \times 2)+1(2 \times 3)=40.
\end{align*}
For $\alpha=-1$, $\mathcal{R}R_{-1}(G)=\frac{3}{1 \times 1}+\frac{9}{1 \times 2}+\frac{3}{1 \times 3}+\frac{1}{2 \times 2}+\frac{1}{2 \times 3}=8.92$.\\
For $\alpha=\frac{1}{2}$, 
\begin{align*}
\mathcal{R}R_{\frac{1}{2}}(G)=&3\sqrt{1 \times 1}+9\sqrt{1 \times 2}+3\sqrt{1 \times 3}+1\sqrt{2 \times 2}+1\sqrt{2 \times 3}=25.37.
\end{align*}
For $\alpha=-\frac{1}{2}$, 
\begin{align*}
\mathcal{R}R_{-\frac{1}{2}}(G)=&\frac{3}{\sqrt{1 \times 1}}+\frac{9}{\sqrt{1 \times 2}}+\frac{3}{\sqrt{1 \times 3}}+\frac{1}{\sqrt{2 \times 2}}+\frac{1}{\sqrt{2 \times 3}}=12.004.
\end{align*}
$\bullet$ Reverse atom bond connectivity index
\begin{align*}
\mathcal{R}ABC(G) & = \displaystyle \sum_{mn \in E(G)} \sqrt{\frac{R_{\psi(m)} + R_{\psi}(n)-2}{R_{\psi(m)} \times R_{\psi}(n)}} \\
& =3\sqrt{\frac{1+1-2}{1 \times 1}}+ 9\sqrt{\frac{1+2-2}{1 \times 2}}+3\sqrt{\frac{1+3-2}{1 \times 3}}+1\sqrt{\frac{2+2-2}{2 \times 2}}+1\sqrt{\frac{2+3-2}{2 \times 3}} \\
& = 10.23.
\end{align*} 
$\bullet$ Reverse geometric arithmetic index
\begin{align*}
\mathcal{R}GA(G)=& \displaystyle \sum_{mn \in E(G)} 2 \frac{\sqrt{R_{\psi(m)} \times R_{\psi}(n)}}{R_{\psi(m)} + R_{\psi}(n)} \\
& = 3\left(\frac{2\sqrt{1 \times 1}}{1+1}\right)+9\left(\frac{2\sqrt{1 \times 2}}{1+2}\right)+3\left(\frac{2\sqrt{1 \times 3}}{1+3}\right) + 1\left(\frac{2\sqrt{2 \times 2}}{2+2}\right)+1\left(\frac{2\sqrt{2 \times 3}}{2+3}\right) \\
 & = 16.06.
\end{align*}
$\bullet$ Reverse first Zagreb index
\begin{align*}
\mathcal{R}M_{1}(G)=& \displaystyle \sum_{mn \in E(G)} \left( R_{\psi(m)} + R_{\psi(n)} \right)\\
& = 3(1+1)+9(1+2)+3(1+3)+1(2+2)+1(2+3)= 54.
\end{align*}
$\bullet$ Reverse second Zagreb index
\begin{align*}
\mathcal{R}M_{2}(G)=& \displaystyle \sum_{mn \in E(G)}  \left(R_{\psi(m)} \times R_{\psi(n)}\right) \\
& = 3(1 \times 1)+9(1\times2)+3(1\times3)+1(2\times2)+1(2\times3)=40.
\end{align*} 
$\bullet$ Reverse first Zagreb co-index
\begin{align*}
\mathcal{R}\overline{M_{1}}(G)=& 2|E(G)|\left(|(V(G)|-1 \right)-\mathcal{R}M_{1}(G) = 2(17)(16-1)-54 = 456.
\end{align*} 
$\bullet$ Reverse second Zagreb co-index
\begin{align*}
\mathcal{R}\overline{M_{2}}(G)=&2|E(G)|^{2}-\frac{1}{2} \mathcal{R}M_{1}(G)-\mathcal{R}M_{2}(G) = 2(17)^2-\frac{54}{2}-40 = 511.
\end{align*}
$\bullet$ Reverse hyper Zagreb index
\begin{align*}
\mathcal{R}HM(G)=& \displaystyle \sum_{mn \in E(G)}  \left(R_{\psi(m)} + R_{\psi(n)}\right)^2 \\
& = 3(1+1)^2+9(1+2)^2+3(1+3)^2+1(2+2)^2+1(2+3)^2= 182.
\end{align*}
$\bullet$ Reverse forgotten index
\begin{align*}
\mathcal{R}F(G)=& \displaystyle \sum_{mn \in E(G)} (R_{\psi(m)})^2 + (R_{\psi(n)})^2\\
& = 3(1^2+1^2)+9(1^2+2^2)+3(1^2+3^2)+1(2^2+2^2)+1(2^2+3^2)= 102.
\end{align*}
$\bullet$ Reverse Balaban index: 
\begin{align*}
\mathcal{R}J(G)=& \frac{m}{m-n+2} \displaystyle \sum_{mn \in E(G)} \frac{1}{\sqrt{R_{\psi(m)} \times R_{\psi(n)}}}\\
& = \left(\frac{17}{17-16+2} \right)\left(\frac{3}{\sqrt{1 \times 1}}+\frac{9}{\sqrt{1 \times 2}}+\frac{3}{\sqrt{1 \times 3}}+\frac{1}{\sqrt{2 \times 2}}+\frac{1}{\sqrt{2 \times 3}}\right) = 68.02.
\end{align*}
$\bullet$ Reverse first multiple Zagreb index
\begin{align*}
\mathcal{R}PM_{1}(G)=&\displaystyle \prod_{mn \in E(G)} \left(R_{\psi(m)} + R_{\psi(n)} \right)\\
& = 3(1+1)\times 9(1+2)\times 3(1+3)\times 1(2+2)\times 1(2+3)= 38880.
\end{align*}
$\bullet$ Reverse second multiple Zagreb index
\begin{align*}
\mathcal{R}PM_{2}(G)=&\displaystyle \prod_{mn \in E(G)} \left(R_{\psi(m)} \times R_{\psi(n)} \right)\\
& = 3(1\times 1)\times 9(1\times 2)\times 3(1\times 3)\times 1(2\times 2)\times 1(2\times 3)= 11664.
\end{align*}
$\bullet$ Reverse first refined Zagreb index
\begin{align*}
\mathcal{R}ReZG_{1}(G)=&\displaystyle \sum_{mn \in E(G)} \frac{R_{\psi(m)} + R_{\psi(n)}}{R_{\psi(m)} \times R_{\psi(n)}}\\
& = 3\left(\frac{1+1}{1 \times 1} \right)+9\left(\frac{1+2}{1 \times 2} \right)+3\left(\frac{2+3}{2 \times 3} \right)+1\left(\frac{2+2}{2 \times 2} \right)+1\left(\frac{2+3}{2 \times 3} \right)= 25.33.
\end{align*}
$\bullet$ Reverse second refined Zagreb index
\begin{align*}
\mathcal{R}ReZG_{2}(G)=&\displaystyle \sum_{mn \in E(G)} \frac{R_{\psi(m)} \times R_{\psi(n)}}{R_{\psi(m)} + R_{\psi(n)}}\\
& = 3\left(\frac{1\times 1}{1 + 1} \right)+9\left(\frac{1\times 2}{1 + 2} \right)+3\left(\frac{2\times3}{2 + 3} \right)+1\left(\frac{2\times2}{2 + 2} \right)+1\left(\frac{2\times3}{2 + 3} \right)= 11.95.
\end{align*}
$\bullet$ Reverse third refined Zagreb index
\begin{align*}
\mathcal{R}ReZG_{3}(G)=&\displaystyle \sum_{mn \in E(G)} \left(R_{\psi(m)} + R_{\psi(n)}\right)\left(R_{\psi(m)} \times R_{\psi(n)} \right)\\
& = 3(2 \times 1)+9(3 \times 2)+3(4 \times 3)+1(4\times 4)+1(5 \times 6)= 142.
\end{align*}    
 
\section{Reverse topological indices of Guadecitabine}
In this section, we find the reverse degree-based topological indices of Guadecitabine and present a graphical comparison. The chemical structure and molecular graph of Guadecitabine are shown in Figure 5.   
\vspace{5mm}
\begin{figure}[h!]
\centering
\includegraphics[width=145mm]{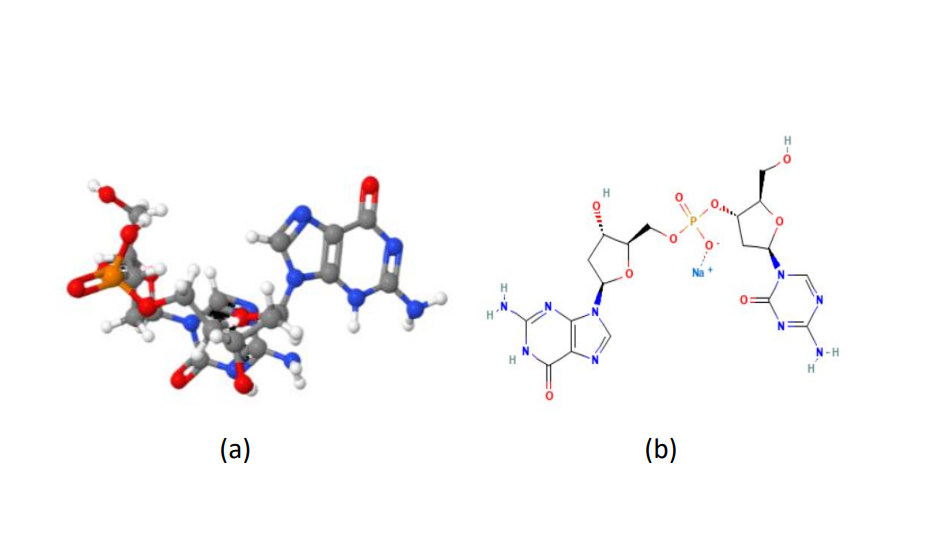}
  \caption{(a) Chemical structure of Guadecitabine. (b) molecular graph of Guadecitabine. }
  \end{figure}  
  
The graph of Guadecitabine with vertices and edges is shown in Figure 6. 
\newpage

\begin{figure}[h!]
\centering
\includegraphics[width=70mm]{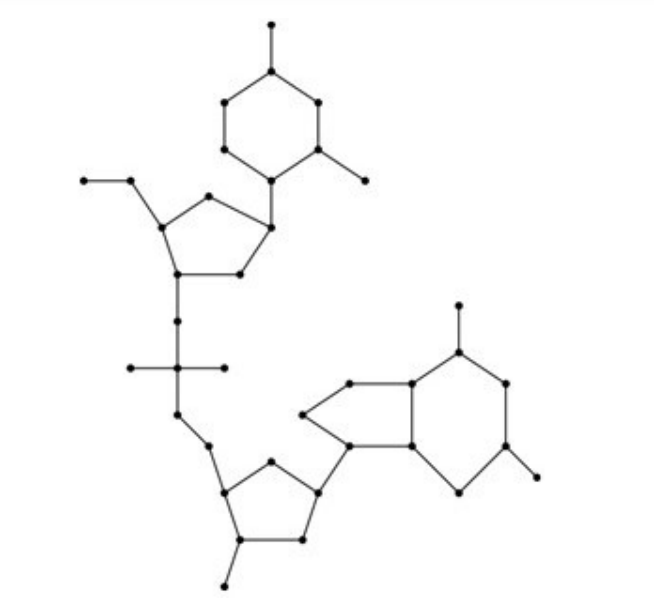}
  \caption{Molecular graph of Guadecitabine with vertices and edges. }
  \end{figure}
  
The number of vertices and edges of the molecular graph of Guadecitabine are 38 and 42, respectively. The edge set of Guadecitabine is partitioned into seven sets based on the reverse degree of end vertices. 

The first edge set of the partition includes 2 edges $mn$, where $R_{\psi(m)}=1$ and $R_{\psi(n)}=3$. The second edge set of the partition includes 2 edges $mn$, where $R_{\psi(m)}=1$ and $R_{\psi(n)}=4$. The third edge set of the partition includes 8 edges $mn$, where $R_{\psi(m)}=2$ and $R_{\psi(n)}=2$. The fourth edge set of the partition includes 21 edges $mn$, where $R_{\psi(m)}=2$ and $R_{\psi(n)}=3$. The fifth edge set of the partition includes 5 edges $mn$, where $R_{\psi(m)}=2$ and $R_{\psi(n)}=4$. The sixth edge set of the partition includes 3 edges $mn$, where $R_{\psi(m)}=3$ and $R_{\psi(n)}=3$. The seventh edge set of the partition includes an edge $mn$, where $R_{\psi(m)}=3$ and $R_{\psi(n)}=4$. 

The maximum degree of the molecular graph of Guadecitabine is 4. The reverse degree-based edge partition of Guadecitabine is given in Table 3. 
\begin{center}
\begin{tabular}{||c c c c c c c c ||} 
 \hline
 $\left(R_{\psi(m)}, R_{\psi(m)}\right)$ & (1,3) & (1,4) & (2,2) & (2,3) & (2,4) & (3,3) & (3,4)  \\ [0.5ex] 
 \hline\hline
 Number of edges & 2 & 2 & 8 & 21 & 5 & 3 & 1 \\ 
 \hline 
\end{tabular}
\end{center}
\begin{center}
\textbf{Table 3}. Edge partition of Guadecitabine.
\end{center}  
The reverse degree-based topological indices of Guadecitabine.\\
$\bullet$ Reverse Randi\'c index
\begin{equation*}
\mathcal{R}R_{\alpha}(G)=\displaystyle \sum_{mn \in E(G)} \left(R_{\psi(m)} \times R_{\psi}(n) \right)^{\alpha}; \alpha=1,-1,\frac{1}{2}, -\frac{1}{2} 
\end{equation*} 
For $\alpha=1$,
\begin{align*}
\mathcal{R}R_{1}(G)=& 2(1 \times 3)+2(1 \times 4)+8(2 \times 2)+21(2 \times 3)+5(2 \times 4)+3(3 \times 3)+1(3 \times 4)=251.
\end{align*}
For $\alpha=-1$, $\mathcal{R}R_{-1}(G)=\frac{2}{1 \times 3}+\frac{2}{1 \times 4}+\frac{8}{2 \times 2}+\frac{21}{2 \times 3}+\frac{5}{2 \times 4}+\frac{3}{3 \times 3}+\frac{1}{3 \times 4}=7.71$.\\
For $\alpha=\frac{1}{2}$, 
\begin{align*}
\mathcal{R}R_{\frac{1}{2}}(G)=&2\sqrt{1 \times 3}+2\sqrt{1 \times 4}+8\sqrt{2 \times 2}+21\sqrt{2 \times 3}+5\sqrt{2 \times 4}+3\sqrt{3 \times 3}+1\sqrt{3 \times 4}=101.51.
\end{align*}
For $\alpha=-\frac{1}{2}$, 
\begin{align*}
\mathcal{R}R_{-\frac{1}{2}}(G)=&\frac{2}{\sqrt{1 \times 3}}+\frac{2}{\sqrt{1 \times 4}}+\frac{8}{\sqrt{2 \times 2}}+\frac{21}{\sqrt{2 \times 3}}+\frac{5}{\sqrt{2 \times 4}}+\frac{3}{\sqrt{3 \times 3}}+\frac{1}{\sqrt{3 \times 4}}=17.78.
\end{align*}
$\bullet$ Reverse atom bond connectivity index
\begin{align*}
\mathcal{R}ABC(G) & = \displaystyle \sum_{mn \in E(G)} \sqrt{\frac{R_{\psi(m)} + R_{\psi}(n)-2}{R_{\psi(m)} \times R_{\psi}(n)}} \\
& =2\sqrt{\frac{1+3-2}{1 \times 3}}+ 2\sqrt{\frac{1+4-2}{1 \times 4}}+8\sqrt{\frac{2+2-2}{2 \times 2}}+21\sqrt{\frac{2+3-2}{2 \times 3}}+5\sqrt{\frac{2+4-2}{2 \times 4}} \\
&+3\sqrt{\frac{3+3-2}{3 \times 3}}+1\sqrt{\frac{3+4-2}{3 \times 4}}\\
& = 30.05.
\end{align*} 
$\bullet$ Reverse geometric arithmetic index
\begin{align*}
\mathcal{R}GA(G)=& \displaystyle \sum_{mn \in E(G)} 2 \frac{\sqrt{R_{\psi(m)} \times R_{\psi}(n)}}{R_{\psi(m)} + R_{\psi}(n)} \\
& = 2\left(\frac{2\sqrt{1 \times 3}}{1+3}\right)+2\left(\frac{2\sqrt{1 \times 4}}{1+4}\right)+8\left(\frac{2\sqrt{2 \times 2}}{2+2}\right) + 21\left(\frac{2\sqrt{2 \times 3}}{2+3}\right)+5\left(\frac{2\sqrt{2 \times 4}}{2+4}\right) \\
& +3\left(\frac{2\sqrt{3 \times 3}}{3+3}\right) +1\left(\frac{2\sqrt{3 \times 4}}{3+4}\right)\\
 & = 40.61.
\end{align*}
$\bullet$ Reverse first Zagreb index
\begin{align*}
\mathcal{R}M_{1}(G)=& \displaystyle \sum_{mn \in E(G)} \left( R_{\psi(m)} + R_{\psi(n)} \right)\\
& = 2(1+3)+2(1+4)+8(2+2)+21(2+3)+5(2+4)+3(3+3)+1(3+4)= 210.
\end{align*}
$\bullet$ Reverse second Zagreb index
\begin{align*}
\mathcal{R}M_{2}(G)=& \displaystyle \sum_{mn \in E(G)}  \left(R_{\psi(m)} \times R_{\psi(n)}\right) \\
& = 2(1 \times 3)+2(1\times 4)+8(2\times 2)+21(2\times 3)+5(2\times 4)+3(3\times 3)+1(3\times 4)=251.
\end{align*} 
$\bullet$ Reverse first Zagreb co-index
\begin{align*}
\mathcal{R}\overline{M_{1}}(G)=& 2|E(G)|\left(|(V(G)|-1 \right)-\mathcal{R}M_{1}(G) = 2(42)(38-1)-210 = 2898.
\end{align*} 
$\bullet$ Reverse second Zagreb co-index
\begin{align*}
\mathcal{R}\overline{M_{2}}(G)=&2|E(G)|^{2}-\frac{1}{2} \mathcal{R}M_{1}(G)-\mathcal{R}M_{2}(G) = 2(42)^2-\frac{210}{2}-251 = 3172.
\end{align*}
$\bullet$ Reverse hyper Zagreb index
\begin{align*}
\mathcal{R}HM(G)=& \displaystyle \sum_{mn \in E(G)}  \left(R_{\psi(m)} + R_{\psi(n)}\right)^2 \\
& = 2(1+3)^2+2(1+4)^2+8(2+2)^2+21(2+3)^2+5(2+4)^2+3(3+3)^2+1(3+4)^2 \\
& = 1072.
\end{align*}
$\bullet$ Reverse forgotten index
\begin{align*}
\mathcal{R}F(G)=& \displaystyle \sum_{mn \in E(G)} (R_{\psi(m)})^2 + (R_{\psi(n)})^2\\
& = 2(1^2+3^2)+2(1^2+4^2)+8(2^2+2^2)+21(2^2+3^2)+5(2^2+4^2)+3(3^2+3^2)+1(3^2+4^2)
\\
& = 570.
\end{align*}
$\bullet$ Reverse Balaban index: 
\begin{align*}
\mathcal{R}J(G)=& \frac{m}{m-n+2} \displaystyle \sum_{mn \in E(G)} \frac{1}{\sqrt{R_{\psi(m)} \times R_{\psi(n)}}}\\
& = \left(\frac{42}{42-38+2} \right)\left(\frac{2}{\sqrt{1 \times 3}}+\frac{2}{\sqrt{1 \times 4}}+\frac{8}{\sqrt{2 \times 2}}+\frac{21}{\sqrt{2 \times 3}}+\frac{5}{\sqrt{2 \times 4}}+\frac{3}{\sqrt{3 \times 3}}+\frac{1}{\sqrt{3 \times 4}}\right) \\
& = 124.46.
\end{align*}
$\bullet$ Reverse first multiple Zagreb index
\begin{align*}
\mathcal{R}PM_{1}(G)=&\displaystyle \prod_{mn \in E(G)} \left(R_{\psi(m)} + R_{\psi(n)} \right)\\
& = 2(1+3)\times 2(1+4)\times 8(2+2)\times 21(2+3)\times 5(2+4) \times 3(3+3) \times 1(3+4) \\
& = 1016064000.
\end{align*}
$\bullet$ Reverse second multiple Zagreb index
\begin{align*}
\mathcal{R}PM_{2}(G)=&\displaystyle \prod_{mn \in E(G)} \left(R_{\psi(m)} \times R_{\psi(n)} \right)\\
& = 2(1 \times 3)\times 2(1 \times 4)\times 8(2\times 2)\times 21(2\times 3)\times 5(2\times 4) \times 3(3\times 3) \times 1(3\times 4) \\
& = 1881169920.
\end{align*}
$\bullet$ Reverse first refined Zagreb index
\begin{align*}
\mathcal{R}ReZG_{1}(G)=&\displaystyle \sum_{mn \in E(G)} \frac{R_{\psi(m)} + R_{\psi(n)}}{R_{\psi(m)} \times R_{\psi(n)}}\\
& = 2\left(\frac{1+3}{1 \times 3} \right)+2\left(\frac{1+4}{1 \times 4} \right)+8\left(\frac{2+2}{2 \times 2} \right)+21\left(\frac{2+3}{2 \times 3} \right)+5\left(\frac{2+4}{2 \times 4} \right)\\
& + 3\left(\frac{3+3}{3 \times 3} \right)+1\left(\frac{3+4}{3 \times 4} \right)\\&= 39.92.
\end{align*}
$\bullet$ Reverse second refined Zagreb index
\begin{align*}
\mathcal{R}ReZG_{2}(G)=&\displaystyle \sum_{mn \in E(G)} \frac{R_{\psi(m)} \times R_{\psi(n)}}{R_{\psi(m)} + R_{\psi(n)}}\\
& = 2\left(\frac{1 \times 3}{1 + 3} \right)+2\left(\frac{1 \times 4}{1 + 4} \right)+8\left(\frac{2 \times 2}{2 + 2} \right)+21\left(\frac{2 \times 3}{2 + 3} \right)+5\left(\frac{2 \times 4}{2 + 4} \right)\\
& + 3\left(\frac{3 \times 3}{3 + 3} \right)+1\left(\frac{3 \times 4}{3 + 4} \right)\\
& =46.26.
\end{align*}
$\bullet$ Reverse third refined Zagreb index
\begin{align*}
\mathcal{R}ReZG_{3}(G)=&\displaystyle \sum_{mn \in E(G)} \left(R_{\psi(m)} + R_{\psi(n)}\right)\left(R_{\psi(m)} \times R_{\psi(n)} \right)\\
& = 2(4 \times 3)+2(5 \times 4)+8(4 \times 4)+21(5\times 6)+5(6 \times 8) +3(6 \times 9)+1(7\times 12) \\
&= 1308.
\end{align*} 

\section{Reverse degree-based quantitative structure-property relationship (QSPR) analysis}
The main objective of this section is to establish a quantitative structure-property relationship (QSPR) between the various topological indices and some physicochemical properties of drugs under study in order to determine the efficacy of the reverse degree-based topological indices. Fifteen reverse degree-based topological indices were used for modeling antiviral activity and eight physicochemical properties, such as boiling point (BP), enthalpy of vaporization (E), flash point (P), molar refraction (MR), polar surface area (PSA), polarizability (P), surface tension (T), and molar volume (MV) of Azacitidine, Decitabine, and Guadecitabine. 

The values for the various physicochemical properties, as presented in Table 4, were collected from ChemSpider. 

\begin{center}
\renewcommand{\arraystretch}{1.75}
\begin{tabular}{|c| c |c |c| c |c |c |c| c |} 
 \hline
 Drugs & BP & E & FP & MR & PSA & P & T & MV \\ [0.5ex] 
 \hline
 Azacitidine & 534.5 & 93.2 & 277 & 51.1 & 141 & 20.3 & 106.8 & 117.1 \\  
 \hline
 Decitabine & 485.8 & 86.6 & 247.6 & 50.2 & 121 & 19.9 & 91.7 & 119.8 \\ 
 \hline
 Guadecitabine & 956.4 & 145.9 & 532.2 & 117.3 & 281 & 46.5 & 131.4 & 248 \\
 \hline 
\end{tabular}
\end{center}
\begin{center}
\textbf{Table 4}. Physicochemical properties of Azacitidine, Decitabine, and Guadecitabine
\end{center} 
By using equations (1-15), the reverse degree-based topological indices of Azacitidine, Decitabine, and Guadecitabine are computed to design the QSPR. 
Table 5 shows the value of reverse degree-based topological indices of Azacitidine, Decitabine, and Guadecitabine. 

\newpage 

In order to form Table 5, the following notations are used.\\
$A=\mathcal{R}R_{1}(G); B=\mathcal{R}R_{-1}(G); C=\mathcal{R}R_{\frac{1}{2}}(G); D=\mathcal{R}R_{-\frac{1}{2}}(G); E=\mathcal{R}ABC(G); F=\mathcal{R}GA(G);\\ G=\mathcal{R}M_{1}(G); H=\mathcal{R}M_{2}(G); I=\mathcal{R}\overline{M_{1}}(G); J=\mathcal{R}\overline{M_{2}}(G); K=\mathcal{R}HM(G); L=\mathcal{R}F(G);\\ M=\mathcal{R}J(G); N=\mathcal{R}PM_{1}(G); O=\mathcal{R}PM_{2}(G); P=\mathcal{R}ReZG_{1}(G); Q=\mathcal{R}ReZG_{2}(G);\\ R=\mathcal{R}ReZG_{3}(G)$.

\begin{center}
\renewcommand{\arraystretch}{1.75}
\begin{tabular}{|c| c| c| c| c| c| c| c| c| c| c|} 
 \hline
Drugs  &A  &B &C &D &E &F &G &H &I &J\\ [0.5ex] 
 \hline
 Azacitidine & 41 & 10.25 & 26.28 & 13.17 & 9.63 & 17.04 & 56 & 41 & 520 & 579 \\ 
 \hline 
 Decitabine & 40 &8.92 &25.37 &12.004 &10.23 &16.06 &54 &40 &456 &511\\
 \hline
Guadecitabine & 251 &7.71 &101.51 &17.78 &30.05 &40.61 &210 &251 &2898 &3172\\ 
 \hline 
\end{tabular}
\end{center}

\begin{center}
\renewcommand{\arraystretch}{1.75}
\begin{tabular}{|c| c| c| c |c |c |c| c| c |} 
 \hline
Drugs  &K  &L &M &N &O &P &Q &R \\ [0.5ex] 
 \hline
 Azacitidine & 188 & 106 & 79.004 & 67200 & 20160 & 27.67 & 12.37 & 146  \\ 
 \hline
 Decitabine & 182 & 102 & 68.02 & 38880 & 11664 & 25.33 & 11.95 & 142 \\
 \hline
Guadecitabine & 1072 & 570 & 124.46 & 1016064000 & 1881169920 & 39.92 & 46.26 & 1308 \\ 
 \hline 
\end{tabular}
\end{center}
\textbf{Table 5}. Reverse degree-based topological indices of Azacitidine, Decitabine, and Guadecitabine. 

\subsection{Regression model} 
Regression analysis is a statistical technique that allows us to explore the relationship between two or more relevant variables. In the context of  cancer treatment, the regression model is employed to identify correlations among the molecular descriptors and physicochemical properties of drugs. The findings of the analysis reveal a strong association between the topological indices and the physicochemical properties of the drugs. 

Linear regression is one of the commonly used statistical techniques in QSPR since it provides a clear and interpretable model. Also, the linear regression method can provide reasonably good results even when the data availability can be limited for certain properties \cite{32}. 

The following equation is used to correlate the various physical properties of Azacitidine, Decitabine, and Guadecitabine.
\begin{equation}
P=A+b[TI]
\end{equation} 
Here, P is a physical property, A is constant, b is the regression coefficient, and TI is a reverse degree-based topological index. 

Table 6 indicates the values of the correlation coefficients ($r$) of the physicochemical properties of Azacitidine, Decitabine, and Guadecitabine with defined reverse degree-based topological indices. 
\vspace{3mm}
\begin{center}
\renewcommand{\arraystretch}{1.75}
\begin{tabular}{|c |c| c| c| c| c |c| c| c |} 
 \hline
Topological index  &BP & E & FP & MR & PSA & P & T & MV \\ [0.5ex] 
 \hline
$\mathcal{R}R_{1}(G)$ & 0.9999 & 0.9952 & 0.9959 & 0.9999 & 0.9938 & 0.9999 & 0.9278 & 0.9995   \\ 
\hline
$\mathcal{R}R_{-1}(G)$ & 0.7990 & 0.7945 & 0.7991 & 0.8458 & 0.7864 & 0.8451 & 0.5920 & 0.8613   \\
\hline
$\mathcal{R}R_{\frac{1}{2}}(G)$ & 0.9964 & 0.9958 & 0.9965 & 0.9999 & 0.9945 & 0.9999 & 0.9301 & 0.9995   \\ 
\hline
$\mathcal{R}R_{-\frac{1}{2}}(G)$ & 0.9952 & 0.9959 & 0.9952 & 0.9837 & 0.9970 & 0.9840 & 0.9811 & 0.9780   \\ 
\hline
$\mathcal{R}ABC(G)$ & 0.9928 & 0.9918 & 0.9928 & 0.9992 & 0.9901 & 0.9992 & 0.9162 & 0.9999   \\
\hline
$\mathcal{R}GA(G)$ & 0.9982 & 0.9977 & 0.9982 & 0.9997 & 0.9968 & 0.9997 & 0.9390 & 0.9985   \\
\hline 
$\mathcal{R}M_{1}(G)$ & 0.9965 & 0.9959 & 0.9965 & 1      & 0.9892 & 0.9999 & 0.9304 & 0.9995   \\ 
\hline
$\mathcal{R}M_{2}(G)$ & 0.9999 & 0.9952 & 0.9959 & 0.9999 & 0.9938 & 0.9999 & 0.9278 & 0.9995  \\
\hline
$\mathcal{R}\overline{M_{1}}(G)$ & 0.9974 & 0.9969 & 0.9974 & 0.9999 & 0.9957 & 0.9999 & 0.9347 & 0.9991   \\ 
\hline
$\mathcal{R}\overline{M_{2}}(G)$ & 0.9974 & 0.9968 & 0.9974 & 0.9999 & 0.9957 & 0.9999 & 0.9345 & 0.9991 \\
\hline
$\mathcal{R}HM(G)$ & 0.9960 & 0.9954 & 0.9961 & 0.9999 & 0.9940 & 0.9999 & 0.9284 & 0.9997  \\ 
\hline
$\mathcal{R}F(G)$ & 0.9962 & 0.9955 & 0.9962 & 0.9999 & 0.9942 & 0.9999 & 0.9290 & 0.9996 \\ 
\hline
$\mathcal{R}J(G)$ & 0.9959 & 0.9965 & 0.9959 & 0.9999 & 0.9975 & 0.9853 & 0.9797 & 0.9795  \\
\hline
$\mathcal{R}PM_{1}(G)$ & 0.9955 & 0.9948 & 0.9955 & 0.9999 & 0.9934 & 0.9999 & 0.9263 & 0.9998  \\ 
\hline
$\mathcal{R}PM_{2}(G)$ & 0.9955 & 0.9948 & 0.9955 & 0.9999 & 0.9934 & 0.9999 & 0.9263 & 0.9998   \\ 
\hline
$\mathcal{R}ReZG_{1}(G)$ & 0.9984 & 0.9988 & 0.9984 & 0.9904 & 0.9993 & 0.9906 & 0.9721 & 0.9859  \\
\hline
$\mathcal{R}ReZG_{2}(G)$ & 0.9965 & 0.9958 & 0.9965 & 0.9999 & 0.9945 & 0.9999 & 0.9302 & 0.9995   \\ 
\hline
$\mathcal{R}ReZG_{3}(G)$ & 0.9958 & 0.9951 & 0.9958 & 0.9999 & 0.9937 & 0.9999 & 0.9274 & 0.9997  \\
 \hline 
\end{tabular}
\end{center}
\vspace{1mm}

\textbf{Table 6}. The correlation between reverse degree-based topological indices and physicochemical properties of Azacitidine, Decitabine, and Guadecitabine. 

\newpage
Figure 7 shows the graphical representation of the correlation of reverse degree-based topological indices with the physical properties of Azacitidine, Decitabine, and Guadecitabine.   
\begin{figure}[h!]
\centering
\includegraphics[width=160mm]{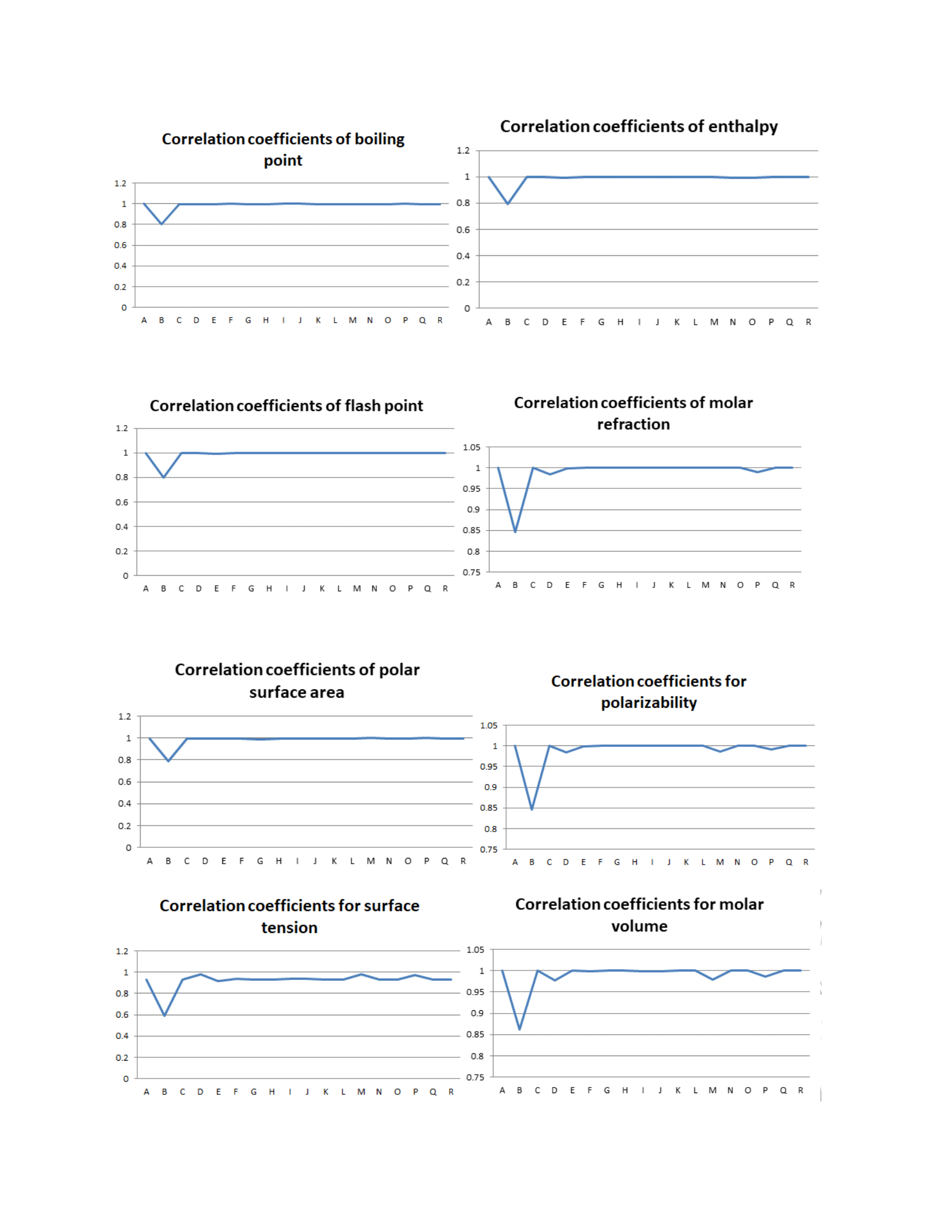}
  \caption{Correlation of physical properties and reverse degree-based topological indices of Azacitidine, Decitabine, and Guadecitabine.}
  \end{figure} 
\newpage

\section{Conclusion}
Table 6 gives the correlation between reverse degree-based topological indices and physicochemical properties of Azacitidine, Decitabine, and Guadecitabine; and graphs of Figure 7 indicate how they correlate. Upon examining correlation coefficients horizontally for boiling point, enthalpy of vaporization, flash point, molar refraction, polar surface area, polarizability, surface tension, and molar volume, it is observed that except for $\mathcal{R}R_{-1}(G)$, every other reverse degree-based topological index gives highest correlation coefficient ($0.9 \leq r \leq 1.0$) for physical properties under consideration. When we look vertically, molar refraction (MR) has a good correlation with $\mathcal{R}R_{-1}(G)$, i.e., ($r=0.8458$). Polarizability (P) and molar volume (MV) also have a good correlation with $\mathcal{R}R_{-1}(G)$, i.e., ($r=0.8451$ and $r=0.8613$). $\mathcal{R}R_{-1}(G)$ does not show good correlation with surface tension (T). From these observations, it may be possible that reverse degree-based topological indices considered above are potential tools for QSPR analysis of cancer drugs.   

\section*{Disclosure statement}
The authors declare that there are no conflicts of interest regarding the publication of this paper.

\makeatletter
\renewcommand{\@biblabel}[1]{[#1]\hfill}
\makeatother


\begin{thebibliography}{99}
\bibitem{1} Syed Ahtsham Ul Haq Bokhary, Adnan, Muhammad Kamran Siddiqui, Murat Cancan, On topological indices and QSPR analysis of drugs used for the treatment of breast cancer, Polycyclic Aromatic Compounds. 42(9) 6233-6253.
\bibitem{2} Leena Rosalind Mary Gnanaraj, Deepa Ganesan, Muhammad Kamran
Siddiqui, Topological indices and QSPR analysis of NSAID drugs, Polycyclic Aromatic Compounds. (2023) 1-17.
\bibitem{3} Deepa Balasubramaniyan,  Natarajan Chidambaram, On some neighbourhood degree-based topological indices with QSPR analysis of asthma drugs, The European Physical Journal Plus. 138, 823 (2023).
\bibitem{4} N. Kansal, P. Garg, O. Singh, Temperature-based topological indices and QSPR Analysis of COVID-19 Drugs, Polycyclic Aromatic Compounds. 43(5) (2023)  4148-4169.
\bibitem{5} M. Kalaimathi, B. J. Balamurugan, Topological indices of molecular graphs of monkeypox drugs for QSPR analysis to predict physicochemical and ADMET properties, International Journal of Quantum Chemistry. (2023)  e27210.
\bibitem{6}  H. Wiener, Structural Determination of Paraffin Boiling Points, Journal of the American Chemical Society. 69 (1) (1947) 17–20.
\bibitem{7} W. Gao, W. Wang, M. R. Farahani, Topological indices study of molecular structure in Anticancer drugs, Journal of Chemistry. 2016 (2016).
\bibitem{8} Tanweer Ul Islam, Z. S. Mufti, A. Ameen, M. N. Aslam, A. Tabraiz, On certain aspects of topological indices, Journal of Chemistry. 2021 (2021).
\bibitem{9} Parvez Ali, S. A. Kirmani, Osamah Al Rugaie, F. Azam, Degree-based topological indices and polynomials of hyaluronic acid-curcumin conjugates,  Saudi Pharm J. 28 (9) (2020) 1093-1100.
\bibitem{10} R. Guha, E. Willighagen, A Survey of Quantitative Descriptions of Molecular Structure,  Curr Top Med Chem. (12) 1946-1956.
\bibitem{11} Adnan, S. A. Ul Haq Bokhary, G. Abbas, T. Iqbal, Degree-Based Topological Indices and QSPR Analysis of Antituberculosis Drugs,  Journal of Chemistry. 2022 (2022).
\bibitem{12} E. J. B. Derissen, J. H. Beijnen, J. H. M. Schellens, Concise drug review: azacitidine and decitabine, Oncologist. 18 (5) (2013) 619-624.
\bibitem{13} R. Daifuku, Pharmacoepigenetics of novel nucleoside DNA methyltransferase inhibitors, Pharmacoepigenetics. 10 (2019) 425-435.
\bibitem{14} V. R. Kulli, Reverse Zagreb and Reverse hyper-Zagreb Indices and Their Polynomials of Rhombus Silicate Networks, Annals of Pure and Applied Mathematics. 16 (1) (2018) 47–51.
\bibitem{15} D. Zhao, Y. M. Chu, M. K. Siddiqui, K. Ali, M. Nasir, M. T. Younas, M. Cancan, On reverse degree based topological indices of polycyclic metal organic network, Polycycl.Aromat.Compd. (2021) 1-18. 
\bibitem{16} C. Y. Jung, M. A. Gondal, N. Ahmad, S. M. Kang, Reverse degree based indices of some nanotubes, J. Discret. Math. Sci. Cryptogr. (22) (2019) 1289-1294.
\bibitem{17} J. Wei, M. Cancan, A. Ur Rehman, M. K. Siddiqui, M. Nasir, M. T. Younas, M. F. Hanif, On topological indices of remdesivir compound used in
treatment of Corona virus (COVID 19), Polycycl. Aromat. Compd. (2021) (1-19).
\bibitem{18} V. Ravi, M. K. Siddiqui, N. Chidambaram, K. Desikan, On topological descriptors and curvilinear regression analysis of antiviral drugs used in COVID-19 treatment, Polycycl. Aromat. Compd. (2021) (1- 14).
\bibitem{19} R. S. Haoer, A. U. R. Virk, Some reverse topological invariants for metal-organic networks, J. Discret. Math. Sci. (24) (2021) 499-510.
\bibitem{20} M. K. Hashmi, F. Chaudhry, A. J. M. Khalaf, M. R. Farhani, Investigation of dendrimer structures by means of reverse atomic bond connectivity index, J. Discret. Math. Sci. Cryptogr. 24 (2021) 473-485. 
\bibitem{21} P. Prasanna, G. B. Gautham, N. Narahari, A. Raghavendra, B. Sooryanarayana, P. Nandini, Reverse topological indices of some molecules in drugs used in the treatment of H1N1, Biointerface Research in Applied Chemistry. 13 (1) (2023) 1-13.
\bibitem{22} M. Chamua, R. Moran, P. Aditya, A. Bharali, M-polynomial and neighborhood M-polynomial of some concise drug structures: Azacitidine, Decitabine, and Guadecitabine, Journal of Molecular Structure. (1263) (2022) 133197.
\bibitem{23} M. Randi\'c, Characterization of Molecular Branching, Journal of the American Chemical Society. 97 (23) (1975) 6609–6615.
\bibitem{24}  E. Estrada, L. Torres, L. Rodriguez, I. Gutman, An Atom-Bond Connectivity Index: Modelling the Enthalpy of Formation of Alkanes, Indian Journal of Chemistry. Sect. A: Inorganic, Physical, Theoretical \& Analytical. 37 (10) (1998) 849–855.
\bibitem{25} D. Vukicevic, B. Furtula, Topological Index Based on the Ratios of Geometrical and Arithmetical Means of End-Vertex Degrees of Edges, Journal of Mathematical Chemistry. 46 (4) (2009) 1369–1376.
\bibitem{26} I. Gutman, N. Trinajstic, Graph Theory and Molecular Orbitals. Total pi-Electron Energy of Alternate Hydrocarbons, Chemical Physics Letters. 17 (4) (1972) 535–548.
\bibitem{27} T. Doslic, Vertex-Weighted Wiener Polynomials for Composite Graphs, Ars Mathematica Contemporanea. 1 (1) (2008) 66–80.
\bibitem{28} I. Gutman, B. Furtula, Z. K. Vukicevic, G. Popivoda, On Zagreb Indices and Coindices, MATCH Communications in Mathematical and in Computer Chemistry. 74 (1) (2015): 5–16.
\bibitem{29} G. H. Shirdel, H. Rezapour, A. M. Sayadi, The hyper-Zagreb Index of Graph Operations, Iranian Journal of Mathematical Chemistry. 4 (2) (2013) 213–220.
\bibitem{30} B. Furtula, I. Gutman, A Forgotten Topological Index, Journal of Mathematical Chemistry. 53 (4) (2015) 1184–1190.
\bibitem{31} P. S. Ranjini, V. Lokesha, A. Usha, Relation between Phenylene and Hexagonal Squeeze Using Harmonic Index, International Journal of Applied Graph Theory. 1 (4) (2013) 116–121.
\bibitem{32} Alan R. Katritzky, Victor S. Lobanov,  Mati Karelson, QSPR: the correlation and quantitative prediction of chemical and physical properties from structure, Chemical Society Reviews. 24 (4) (1995) 279-287.


\end{thebibliography}
\end{document}